# Experimental demonstrations of coherence de Broglie waves using sub-Poisson distributed coherent photon pairs

S. Kim and Byoung S. Ham
School of Electrical Engineering and Computer Science, Gwangju Institute of Science and Technology
123 Chumdangwagi-ro, Buk-gu, Gwangju 61005, S. Korea
(Submitted on July 22, 2021; bham@gist.ac.kr)

**Abstract:** Recently, a new interpretation of quantum mechanics has been developed for the wave nature of a photon, where determinacy in quantum correlations becomes an inherent property without the violation of quantum mechanics. Here, we experimentally demonstrate a direct proof of the wave natures of quantum correlation for the so-called coherence de Broglie waves (CBWs) using sub-Poisson distributed coherent photon pairs obtained from an attenuated laser. The observed experimental data coincides with the analytic solutions and the numerical calculations. Thus, the CBWs pave a road toward deterministic and macroscopic quantum technologies for such as quantum metrology, quantum sensing, and even quantum communications, that are otherwise heavily limited due to the microscopic non-determinacy of the particle nature-based quantum mechanics.

**Introduction**

Photonic de Broglie waves (PBWs) based on two-mode entangled photon pairs have witnessed quantum sensing and quantum metrology to break the classical limits of sensitivity and resolution[1-4]. Although PBWs have been intensively studied for quantum metrology[5-13] and quantum sensing[14-24] over the last few decades, their implementations are still severely limited due to the difficulties and inefficiency of higher-order NOON state generations[4]. Recently a coherence version of PBWs, the so-called coherence de Broglie waves (CBWs), have been proposed in a macroscopic regime of asymmetrically coupled Mach-Zehnder interferometers (MZIs)[25]. Unlike PBWs based on the particle nature of quantum mechanics, CBWs originate from the wave nature of a photon in a coupled MZI system via phase basis superposition. Experimental demonstrations have also been performed to prove the nonclassical nature of CBWs using the frequency modulated classical light of a laser[26]. Here, a quantum version of CBW observations is presented using nearly sub-Poisson distributed coherent photons achieved from an attenuated laser. To understand the quantum superposition-based nonclassical features of CBWs, various violation tests and numerical simulations are conducted, where asymmetrical phase coupling between paired MZIs is the key control parameter for the CBWs.

Very recently, coherence interpretations of quantum mechanics have been performed for fundamental quantum features such as anticorrelation on a beam splitter (BS)[27] and unconditional security in classical key distributions (USCKD)[28], in which quantum superposition between phase bases of an MZI plays a major role for the generation of nonclassical phenomena. The core concept of quantum entanglement has been viewed for a nonlocal correlation that cannot be achieved by classical means[29,30]. According to recent studies[25-28], however, quantum features can also be viewed as coherence optics without violating quantum mechanics according to the complementarity theory[31]. In other words, the conventional nonlclassical features are newly interpreted as a special case of coherence optics for maximal coherence[32]. As a result, an interferometric system such as an MZI has been applied for a quantum device via phase basis superposition[25-28,32], where the new interpretation of quantum correlation is totally different from the conventional approach based on the particle nature of a photon. The wave nature-based new interpretation of quantum correlation reveals phase basis-determined macroscopic quantum properties. Due to the phase controllability, macroscopic quantum entanglement generation becomes an inherent feature of the present CBWs for potential applications of quantum technologies.

In a single MZI, two orthonormal phase bases are 0 and $\pi$, resulting in strong correlation between the output paths and the phase bases[27]. This phase-path correlation has no relation with the photon numbers determining Fock states and shows the same quantum entanglement relation between phase-path correlations. To fulfill the quantum entanglement in an MZI, basis choice (or measurement) randomness may be satisfied between the phase-path correlations. This is the macroscopic randomness in the new interpretation of quantum mechanics, where it can be provided either in a doubly-coupled MZI in the name of quantum lasers[32] or a triply-coupled MZI in the name of CBWs[25]. On the contrary of the observed CBWs using frequency modulated commercial laser light without any quantum treatments, the present CBW demonstrations use quantum light of nearly sub-Poisson distributed photon pairs achieved from intensity attenuation of a commercial laser. According to the wave nature of a photon in quantum mechanics, the photon characteristics are related with phase (coherence) rather than energy (photon number), where the phase and energy of a photon are mutually exclusive. In CBWs, the phase property can also be quantized via quantum superposition between phase bases in a coupled interferometric system (discussed elsewhere). This is the critical difference from classical (coherence) optics as well as conventional particle nature-based quantum optics.

*Experimental setup*

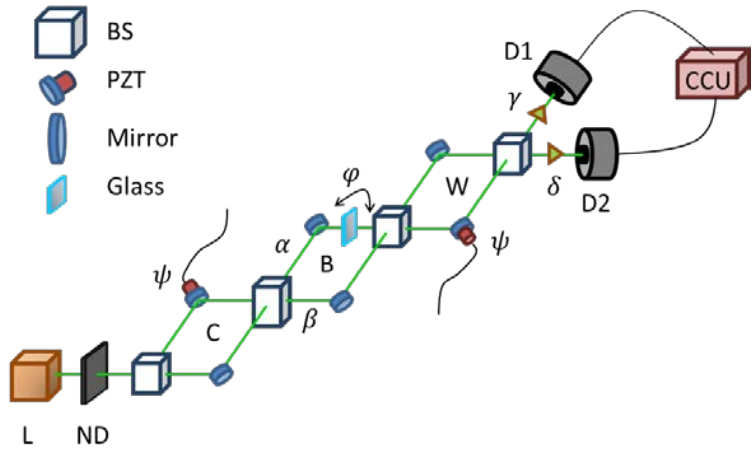

**Fig. 1| An experimental set-up for CBWs.** L: 532 nm cw laser, BS: unpolarized 50/50 beam splitter, ND: neutral density filters, PZT: piezo-electric transducer, D: single photon detector, CCU: coincidence counting unit. C, B, and W represent phase-coupled MZIs. $\Psi$s are synchronized by a function generator. α, β, γ, and δ stand for output photons or fields of the corresponding MZIs.

Figure 1 shows a schematic of a CBW, where C, B, and W stand for individually phase-controlled MZIs. For the asymmetric phase relation between them, the C and W MZIs are coupled via the B MZI. For the theoretical point of view, the basic building block for the phase basis superposition is composed of C and B MZIs, where the following W is supposed to have another B MZI. Here, the fundamental requirement for the CBW is an asymmetric relation between consecutive CBW modules to have an alternative π phase shift. The missing B MZI after the W MZI in Fig. 1 has no influence in the measurement results, where the photon γ is supposed to have a different phase from the photon δ when detected, and vice versa. The nonclassical input photon $E_0$ entering the CBW interferometric system is achieved by optical attenuation of a 532 nm continuous-wave (cw) laser (Coherent, V-10) with a 5 MHz spectral linewidth by using neutral density filters (ND). The sub-Poisson distributed photon characteristics have been theoretically and experimentally confirmed with the counted mean photon number of $\langle n \rangle \sim 0.04$ (see Section 1 of the Supplementary Information).

      The main phase controllers ψs in Fig. 1 are synchronized by the same function generator (Tektronix AFG3102)-controlled piezo-electric transducers (PZTs; Thorlabs, POLARIS-K2S3P). Here, the intermediate MZI B plays an important role for the phase basis superposition between the coupled MZIs of C and W[25]. Contrary to PBWs based on the pre-entangled photons from the second-order nonlinear effects of spontaneous parametric down conversion processes (SPDC), the entanglement of CBWs in Fig. 1 is based on cascaded MZIs of C and W for

antisymmetric phase basis superposition via the control MZI B. As a consequence, higher-order CBWs can be generated deterministically by using linear superposition of the CBW modules[25]. The reason of using quantum particles of single photons in the present demonstration is to demonstrate the quantumness of CBWs for its wave nature.

*Review of the CBW Theory*

For theoretical analyses, the interference fringes of the individual output photons (or fields), α, β, γ, and δ in Fig. 1 are driven by the use of pure coherence optics as follows. Using typical MZI matrix representations[25], the output fields' amplitudes, $E_\alpha$ and $E_\beta$, from the first MZI C are obtained:

$$\begin{bmatrix} E_\alpha \\ E_\beta \end{bmatrix} = [MZI]_C \begin{bmatrix} E_0 \\ 0 \end{bmatrix},$$

$$= \frac{1}{2}\begin{bmatrix} 1 - e^{i\psi} & i(1 + e^{i\psi}) \\ i(1 + e^{i\psi}) & -(1 - e^{i\psi}) \end{bmatrix}\begin{bmatrix} E_0 \\ 0 \end{bmatrix}, \quad (1)$$

where $[MZI]_C = [BS][\psi][BS]$, $[\psi] = \begin{bmatrix} 1 & 0 \\ 0 & e^{i\psi} \end{bmatrix}$, and $[BS] = \frac{1}{\sqrt{2}}\begin{bmatrix} 1 & i \\ i & 1 \end{bmatrix}$. Thus, the corresponding output intensities $I_\alpha$ and $I_\beta$ are, respectively:

$$I_\alpha = \frac{I_0}{2}(1 - \cos\psi), \quad (2)$$

$$I_\beta = \frac{I_0}{2}(1 + \cos\psi), \quad (3)$$

where $I_0 = E_0 E_0^*$. Equations (2) and (3) represent the phase basis-dependent MZI path directionality as a direct result of coherence optics as well as the classical bound of diffraction limit, $\lambda_0/2$, where $\lambda_0$ is the wavelength of the input field $E_0$. Here, the intensity modulation period of $I_\alpha$ and $I_\beta$ is $2\pi$ or $\lambda_0$, where the corresponding phase basis to the output paths of MZI is $\psi \in \{0, \pi\}$. As is well known by Born's rule, there is no distinction between a quantum particle of a single photon[33] and a classical wave of a coherent laser light for interferometric results[34], where quantum mechanics is governed by the square law of amplitude probability in measurements[35]. In other words, the output field's characteristics of a single MZI have no difference between a single photon and a wave (of a typical laser light) as an input[35].

From equations (2) and (3), the final outputs of CBWs for Fig. 1 are driven as follows:

$$\begin{bmatrix} E_\gamma \\ E_\delta \end{bmatrix} = [MZI]_W[\varphi]\begin{bmatrix} E_\alpha \\ E_\beta \end{bmatrix}$$

$$= \frac{1}{4}\begin{bmatrix} -(1 - e^{i\psi})(1 - e^{i\psi}) - e^{i\varphi}(1 + e^{i\psi})(1 + e^{i\psi}) & -i(1 + e^{i\psi})(1 - e^{i\psi}) + ie^{i\varphi}(1 + e^{i\psi})(1 - e^{i\psi}) \\ i(1 + e^{i\psi})(1 - e^{i\psi}) + ie^{i\varphi}(1 + e^{i\psi})(1 - e^{i\psi}) & -(1 + e^{i\psi})(1 + e^{i\psi}) - e^{i\varphi}(1 - e^{-i\psi})(1 - e^{i\psi}) \end{bmatrix}\begin{bmatrix} E_0 \\ 0 \end{bmatrix}, \quad (4)$$

where $[MZI]_W = [BS][\psi'][BS]$, $[\psi'] = \begin{bmatrix} e^{i\psi} & 0 \\ 0 & 1 \end{bmatrix}$, and $[\varphi] = \begin{bmatrix} e^{i\varphi} & 0 \\ 0 & 1 \end{bmatrix}$.

(i) For $\varphi = 0$ (Asymmetric condition)

For CBWs with $\varphi = 0$ (or $2n\pi$), equation (4) becomes:

$$\begin{bmatrix} E_\gamma \\ E_\delta \end{bmatrix} = -\frac{1}{2}\begin{bmatrix} 1 + e^{i2\psi} & i(1 - e^{i2\psi}) \\ -i(1 - e^{i2\psi}) & 1 + e^{i2\psi} \end{bmatrix}\begin{bmatrix} E_0 \\ 0 \end{bmatrix}, \quad (5)$$

where the corresponding final output intensities of CBWs are as follows:

$$I_\gamma = \frac{I_0}{2}(1 + \cos 2\psi), \quad (6)$$

$$I_\delta = \frac{I_0}{2}(1 - cos2\psi). \tag{7}$$

Compared with equations (2) and (3) governed by classical optics, the phase resolution of individual outputs in equations (6) and (7) is doubly enhanced to $\pi/2$ or $\lambda_0/4$. This effect is the same as with a half-cut wavelength $\lambda_0/2$ in a single MZI, i.e., $\lambda_{CBW} = \lambda_0/2n$ (n=1), which corresponds to a PBW at $\lambda_B = \lambda_0/2N$ (N=2). The factor 2 enhancement in the PBW is due to coincidence detection. Here, n in $\lambda_{CBW}$ represents the number of the basic building blocks composed of C and B MZIs in Fig. 1. Thus, equations (6) and (7) successfully derive the quantum nature of CBWs, where the n corresponds to the entangled photon number N in PBWs. The resulting $\lambda_{CBW}$ violates the classical physics governed by the diffraction limit of the Rayleigh criterion at $\lambda_0/2$ [36]. The control phase basis of φ in the B MZI plays a crucial role for the creation of quantum features between C and W. For this, the CBW for a symmetric condition with an opposite φ basis is analyzed below.

(ii) For $\varphi = \pi$ (Symmetric condition)

For the opposite case of (i), a symmetrically coupled MZI for the C and W MZIs is configured with $\varphi = \pi$ for the B MZI in Fig. 1. As a result, equation (4) becomes:

$$\begin{bmatrix} E_\gamma \\ E_\delta \end{bmatrix} = e^{i\psi} \begin{bmatrix} 1 & 0 \\ 0 & -1 \end{bmatrix} \begin{bmatrix} E_0 \\ 0 \end{bmatrix}. \tag{8}$$

The corresponding output intensities of equation (8) are $I_\gamma = I_0$ and $I_\delta = 0$ (see Fig. 3). This symmetric CBW with $\varphi = \pi$ is for USCKD[28], where the double unitary transformations in a folded MZI scheme of Fig. 1 results in an identical matrix relation between the input and outputs via deterministic randomness:

$$I_\gamma = I_0, \tag{9}$$

$$I_\delta = 0. \tag{10}$$

In equations (9) and (10), the output intensities can be swapped if the choice of ψ basis is opposite for both the C and W MZIs. As proposed[28] and experimentally demonstrated[37], USCKD is an extreme feature of CBWs, where the anti-eavesdropping is due to MZI phase-basis randomness[28,37] rather than the no cloning theorem[38].

*Experimental results: CBW*

Figure 2 shows experimental demonstrations of the CBWs in Fig. 1 for equations (1) and (2) using sub-Poisson distributed coherent photons, where the phase control of ψ is conducted by a pair of synchronized piezo-electric transducers (PZTs) controlled by a function generator-triggered linear voltage ramp. The linear voltage ramp is a right triangle shape of 0 to 100 V. The full PZT scan time for this voltage ramp is 500 s, where the collection time of each data point is 0.1 s, resulting in 5000 data points for a full scan, as shown in Fig. 2. For a single MZI, the total number of resulting coincidence fringes is 21, where Fig. 2a shows a part for 200 s scanning. The intensity fluctuations of the fringe maxima of the coincidence detection in Fig. 2 is both due to air turbulence as well as the photons' miss-alignment on a beam splitter (BS).

Figure 2a shows typical MZI outputs from the C MZI in Fig. 1 as a classical limit of diffraction optics, whose fringe modulation period is a full wavelength $\lambda_0$ of the 532 nm laser and the fringe visibility is $99.5 \pm 0.04$ %. Due to the 'AND' logic gate operation for the coincidence measurements ($R_{\alpha\beta}$) by the coincidence counting unit (CCU) between the two outputs α and β, the fringe modulation is doubled as shown by the blue curve compared with the individual ones (see the red and black curves). The measured coincidence fringe visibility is $98.3 \pm 0.15$ %. The coincidence count rate is ~1% of the single photon count rates for the mean photon number of $\langle n \rangle \sim 0.04$ (see Section 1 of the Supplementary Information). In other words, three or more bunched photons are neglected with ~1 % statistical error.

Figure 2b is for the experimental demonstration of the CBWs in Fig. 1 for the case of the asymmetric coupling with $\varphi = 0$, where the corresponding fringe modulations are doubled compared to those in Fig. 2a, respectively. As confined by the diffraction limit, such an increased fringe modulation cannot be obtained by classical physics, demonstrating a direct proof of the quantum correlation in CBWs. Unlike nondeterministic PBWs based on the particle nature of quantum mechanics, Fig. 2b is obviously deterministic by the choice of a phase basis and the number of CBW modules. The individual fringe visibilities of the output photons γ and δ are respectively 99.6 ± 0.03 % and 99.6 ± 0.02 %. The coincidence $(R_{\gamma\delta})$ fringe visibility of $V_{CBW} = 98.5 \pm 0.39$ % far greater than the classical bound of $V_{CBW} = 70.7$ % also demonstrates the quantum feature of CBWs. Thus, sub-Poisson distributed photon-based CBWs are successfully demonstrated for quantum correlation using attenuated coherent photons.

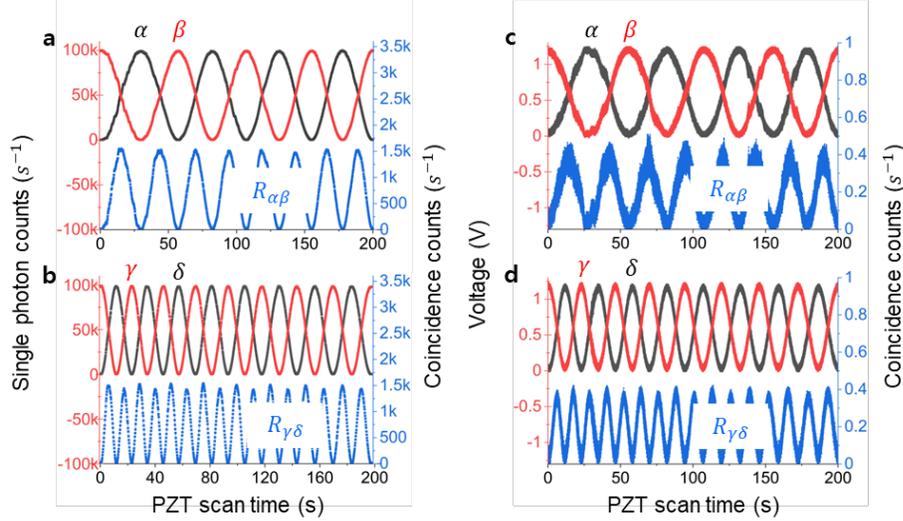

**Fig. 2| Experimental demonstrations of CBW for $\varphi = 0$ in Fig. 1. a**, Individual and coincidence (blue) detection count rates for α (black) and β (red). **b**, Individual and coincidence (blue) detection count rates for γ (black) and δ (red). **c,d**, Corresponding cw cases for **a** and **b**, respectively. PZT scan time relates to a voltage ramp from 0 V to 100 V. Each data acquisition time is for 0.1 s. In **a**, the detectors of D1 and D2 are moved to detect the signals of α and β, respectively. For coincidence counts (blue), see the right vertical axis.

The fringe pattern of the 1st intensity correlations in Fig. 2b (see the red and black curves) is due to the phase basis deterministic self-interference of a single photon on the BS, whereas phase basis randomness results in uniform intensities[32]. The observed nonclassical feature of an entangled photon pair of CBWs in Fig. 2b is completely different from that in PBWs. Unlike all other (conventional) quantum features based on SPDC-generated entangled photon pairs, all photon pairs in Fig. 1 are definitely coherent within the spectral bandwidth of 5 MHz. Thus, the violation of randomness in the phase basis choice in Fig. 2 is a unique feature of CBWs.

Figures 2c and 2d show for typical laser lights to demonstrate the wave nature as well as Born's rule[34]. For this, the input field intensity of $E_0$ is set at 50 μW. As experimentally demonstrated using frequency modulated light[26], the observed identical features are not surprising at all due to the inherent property of the wave nature of a single photon. As mentioned by Dirac[35], this result proves the self-interference of a single photon, where photons do not interfere with others. In other words, MZIs do not distinguish whether the input is a single photon or a classical light. Thus, the wave nature-based quantum features of CBWs are successfully demonstrated for unprecedented quantum features using nonclassical coherent photons, resulting in deterministic and macroscopic properties of CBWs. Understanding the (relative) phase information of the photon pairs, thus, should lift off the mysterious and spooky quantum correlations in conventional quantum mechanics as mentioned by Einstein[39].

Figure 3 shows some violations of CBWs with different phase basis choice of φ. For the symmetric MZI configuration with $\varphi = \pi$, the CBW turns out to be USCKD as shown in Fig. 3a[28,37]. USCKD has been studied for its inherent property of channel eavesdropping randomness, satisfying unconditional security even for classical key distribution. As analyzed in equation (4), the output characteristics of USCKD are also pre-determined. Unlike typical quantum key distribution (QKD) protocols based on the particle nature of a photon[38], Fig. 3a reveals the wave version of QKD, corresponding to the relation between PBWs and CBWs. Thus, both deterministic and macroscopic properties can be applied for potential applications of future secured classical communications compatible with current optical technologies.

Figure 3b is for an example of uncontrolled phase basis with $\varphi = \pi/2$ to show an error or violation of CBWs at $\lambda_{CBW}$, resulting in the loss of quantumness (discussed in Fig. 4). Figures 3c and 3d are corresponding cases of cw laser light as an input $E_0$. As demonstrated in Figs. 2c and 2d, Figs. 3c and 3d also show the same feature of the wave nature.

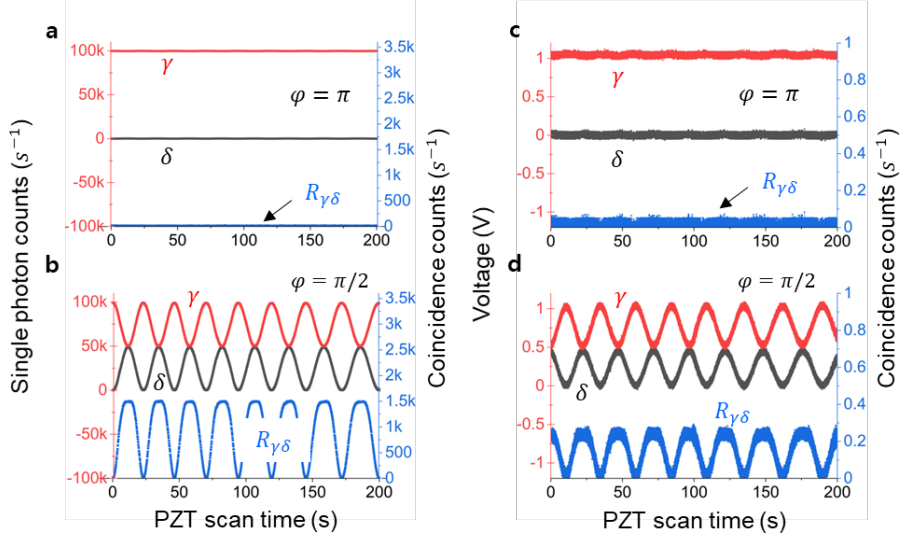

**Fig. 3| Analysis of CBWs. a,b,** CBW for $\varphi = \pi$ and $\varphi = \pi/2$ for Fig. 1. **c,d,** Corresponding cw cases for **a** and **b**, respectively.

Figure 4 represents numerical simulations for a full scale of CBWs using equations (1) and (4). Figures 4a-4d are for the CBWs (n=2) observed in Figs. 2 and 3, while Figs. 4e and 4f are for n=3. The red (blue) curve in Fig. 4c is the detail of Figs. 4a (4b) for the present asymmetric condition with $\varphi = 0$, corresponding to Fig. 2b. The dotted curve in Fig. 4c is for the coincidence measurements, corresponding to the blue curve in Fig. 2b. The individual output data of $I_\gamma$ and $I_\delta$ are normalized, otherwise they double in intensity due to the bipartite entities of inputs. The dotted curve in Fig. 4c shows the same modulation as the PBW for N=2 in ref. 3, resulting in the same doubly enhanced phase resolution of a nonclassical feature. As shown in the experiments in Fig. 2b, however, CBWs also show a fringe pattern for individual outputs due to coherence optics. Here, it should be noted that the uniform individual detection in PBWs is due to coherence washout among SPDC generated photon pairs. Figure 4d corresponds to Fig. 3 for both $\varphi = \pi/2$ (solid) and $\varphi = \pi$ (dotted), representing a control phase $\varphi$ choice for the CBW error and USCKD, respectively. Thus, the observed experimental demonstrations are strongly supported by the CBW theory as well as numerical simulations.

Figures 4e and 4f are for a higher-order CBW with n=3. For this, the three CBW modules composed of C and B in Fig. 1 are combined in series. The corresponding matrix representation for this is as follows:

$$\begin{bmatrix} E_\gamma \\ E_\delta \end{bmatrix} = [\varphi][MZI]_C[\varphi][MZI]_W[\varphi][MZI]_C \begin{bmatrix} E_0 \\ 0 \end{bmatrix}. \tag{11}$$

Equation (11) is numerically calculated in Figs. 4e and 4f, where the fringe of individual output channels (solid curves) in Fig. 4f is tripled compared to the classical limit (dotted curve). The green box in Fig. 4f is for the discussion of phase sensitivity governed by the Heisenberg limit, which is also triply enhanced (see Discussion). Unlike the nondeterministic and mysterious quantum features studied over the last several decades, the present

CBWs can be clearly understood in terms of the phase basis superposition between the paired entities without violation of quantum mechanics[40]. Thus, the phase basis choice in Fig. 1 for CBWs determines its quantum characteristics.

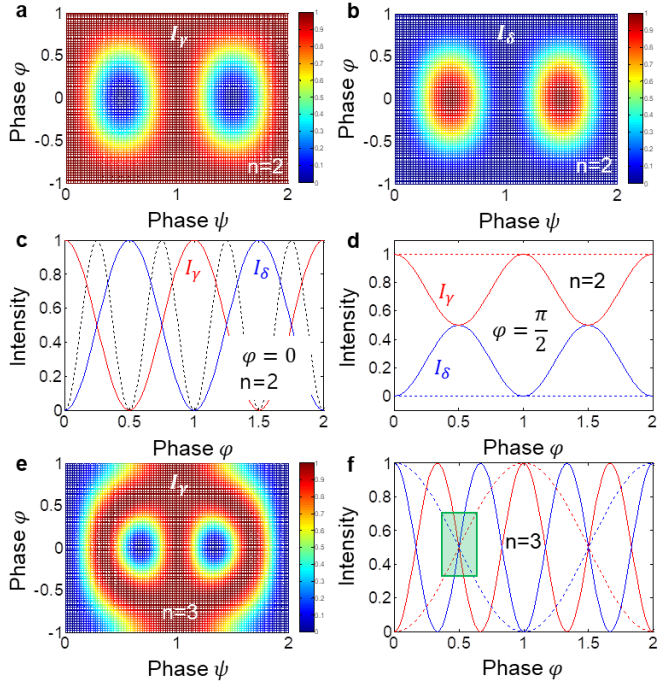

**Fig. 4| Numerical calculations of CBWs. a-d**, CBWs for n=2. **e**, $I_\gamma$ for n=3. **f**, Output intensities for n=1 (dotted) and n=3 (solids). Red (Blue) solid: $I_\gamma$ ($I_\delta$). Blue (red) dotted: $I_\alpha$ ($I_\beta$). Green box: phase sensitivity (see Discussion).

*Discussion*

The origin of the observed nonclassical features of CBWs in Fig. 2 is due to the quantum superposition of phase bases in an asymmetrically coupled MZI system. The phase basis-determined output photons from the first MZI C in Fig. 1, $E_\alpha$ and $E_\beta$ are fed into the synchronously coupled MZI W via a control MZI B, resulting in the outputs $E_\gamma$ and $E_\delta$ via phase basis superposition. Each final output results in doubled fringe modulation due to phase-basis quantization determined by the CBW at $\lambda_{CBW} = \lambda_0/2n$, where n is the number of basic CBW module composed of C and B MZIs. This corresponds to the energy quantization of an entangled photon pair in PBWs[8].

The increased fringe modulation of the final outputs in Fig. 4f results in the n-times increased phase sensitivity $\Delta\varphi_{CBW}$ compared with the classical bound in a typical MZI output, where the classical sensitivity is $\Delta\varphi = \eta/I_\alpha$ (see Section 2 of the Supplementary Information):

$$\Delta\varphi_{CBW} = \frac{\Delta\varphi}{n}, \tag{12}$$

where η is the effective maximum slope of the output intensity difference $\Delta I_{\alpha\beta}$ ($= I_\alpha cos\varphi$) at $\varphi = \pi/2$ (see the green box in Fig. 4f). For even n, however, the maximum slope occurs at $\varphi = \pi/2n$ due to the phase basis quantization (see Fig. 4c). Unlike the extremely low η for higher Ns in PBWs, CBWs result in near unity regardless of n due to the coherence optics of the MZIs. As n in the CBWs increases, not only the phase resolution, but also the phase sensitivity of CBWs increases by the same ratio of n. This is the unprecedented quantum feature of CBWs as demonstrated in Figs. 2 and 4.

Photon numbers in each path of the MZI have nothing to do with this new type of quantum correlation in CBWs. As observed in Figs. 2 and 3, the coupled quantum superposition is only dependent upon the binary basis of coupled MZIs. Thus, CBWs have great advantages in quantum sensing and quantum metrology over PBWs with unlimited light intensity such as squeezed light applied for gravitational wave detection[41]. Regarding inertial navigation, a quantum Sagnac interferometer would be a good candidate for potential applications of the present CBWs in a rotating frame of an optical cavity[42], where Sagnac interferometers have been used for ultrahigh precision/positioning systems of unmanned vehicles, drones, and submarines[43].

*Conclusion*

Nonclassical features of the coherence de Broglie waves (CBWs) were experimentally demonstrated for the first time using nearly sub-Poisson distributed coherent photon pairs in an asymmetrically coupled MZI system, where CBWs originate in the wave nature of quantum mechanics via phase basis superposition. Unlike the conventional understanding of quantum features based on the particle nature of a photon, CBWs have a great benefit in potential applications of quantum sensing and quantum metrology for both enhanced phase resolution and sensitivity. Moreover, the observed CBWs also demonstrated macroscopic and deterministic quantum features. Thus, the observed CBWs open the door to on-demand quantum technologies not only for quantum sensing and quantum metrology but also for quantum communications based on higher-order entangled photon pairs, overcoming the critical limitations of nondeterministic and impractical N00N-based quantum technologies. The higher-order CBWs also provide significant benefits in quantum lithography, quantum imaging, quantum inertial navigation, and gravitational wave detections using CBW lights due to their near perfect visibility even for higher-order CBWs.

**Methods**

The laser L in Fig. 1 is a commercial 532 nm cw laser (Coherent, V-10), whose spectral linewidth is as narrow as 5 MHz. The corresponding coherent length is 60 m, which is long enough for the coherence condition of the MZI system of CBW. The dimension of each MZI in Fig. 1 is ~10x10 $cm^2$. The MZIs in Fig. 1 is enclosed by a black cotton box to minimize measurement errors from air turbulence and stray light. The phase stability of the MZI system persists ~30 minutes within a few per cent of intensity fluctuation. Avalanche photodiodes (Hamamatsu, C12703) are used to record the classical case of CBWs of Figs. 2c and 2d, where the data in Figs. 2c and 2d are directly from the record in an oscilloscope (Yokogawa, DL9040; 500 MHz).

For the attenuation of the 532 nm laser, ND filters are used for optical density (OD) of 13, resulting in the mean photon number of $\langle n \rangle = 0.04$. For the analyses of attenuated photon statistics, see Section 1 of the Supplementary Information. The photon characteristics used for experiments in Fig. 2 were well satisfied for the nearly antibunched single photon stream. For CBW in Fig. 2, however, only doubly bunched photons are post-selected for the coincidence detection measurements, where the doubly bunched to single photons is ~1 % for the mean photon number of $\langle n \rangle \sim 0.04$. All single photon detections were made by a pair of single photon counting modules (SPCMs: Excelitas, AQRH-SPCM-15).

The total data point measured in each fringe is 1,000. The PZT scan times for each fringe in D1, D2 and R12 are $49.8 \pm 4.6$ s, $49.8 \pm 4.9$ s and $23.3 \pm 1$ s in Fig. 2a, and $27 \pm 2.7$ s, $27 \pm 2.9$ s and $12 \pm 1.4$ s in Fig. 2b, respectively. The visibility with standard deviation in Fig. 2b is calculated by averaging measured values for all fringes between the maxima and minima. For the maxima and minima, a homemade Matlab program is used.

For the control phase $\varphi$ in Fig. 1, a thin glass plate whose thickness is 1 mm was used for the transmitted light, where rotation angle $\theta$ from the normal incidence of a photon induces path-difference $\Delta L$, $\Delta L = L_0 \left( \frac{n}{\sqrt{1-sin^2\theta}} - 1 \right)$, where n is the refractive index of the glass. At $\theta = 45°$, $\Delta L \sim 6$ μm (or ~12 fringes) per degree (see Section 3 of the Supplementary Information). Thus, a fine tuning of the glass rotation for the basis basis of $\varphi$ were manually carried out easily by using a micrometer, where the $\varphi$ basis was examined by measuring the outputs of MZI B with blocking one path of MZI C. $I_\alpha$ and $I_\beta$ according to equations (2) and (3). For $\varphi = \pi/2$, $I_{cu} = I_{cl}$ was achieved, where cu (cl) represents MZI C's upper (lower) path. For $\varphi = \pi$ (0), $I_{cu} = 0$ ($I_0$) or vice versa.

**Acknowledgment:** The author acknowledges that the present work was supported by a GRI grant funded by GIST in 2021.

**Author contribution**

B.S.H. conceived the idea, analyzed the data with numerical calculations, and wrote the manuscript. S.K. conducted the experiments and analyzed the related data. Correspondence and request of materials should be addressed to BSH (email: bham@gist.ac.kr).

**Competing interests**

The author declares no competing (both financial and non-financial) interests.

Supplementary information is available in the online version of the paper. Reprints and permissions information is available online at www.nature.com/reprints.